\documentclass[12pt]{article}
\usepackage{amsmath,epsfig,graphicx,amssymb}

\newcommand{\beq}{\begin{equation}}
\newcommand{\eeq}{\end{equation}}
\newcommand{\ben}{\begin{eqnarray}}
\newcommand{\een}{\end{eqnarray}}
\newcommand{\bes}{\begin{subequations}}
\newcommand{\ees}{\end{subequations}}
\newcommand{\bFig}{\begin{figure}}
\newcommand{\eFig}{\end{figure}}

\date{}
\begin{document}

\title{Novel States of Classical Light and Noncontextuality}
\author{Partha Ghose\footnote{partha.ghose@gmail.com} \\
Centre for Astroparticle Physics and Space Science (CAPSS),\\Bose Institute, \\ Block EN, Sector V, Salt Lake, Kolkata 700 091, India, \\and\\ Anirban Mukherjee,\\Indian Institute of Science Education \& Research,\\Mohanpur Campus, West Bengal 741252.}
\maketitle
PACS nos. 42.25.-p, 42.25.Ja, 42.25.Kb, 03.65.Ud
\vskip 0.2in
Keywords: classical light, polarization, entanglement, nonseparability, noncontextuality
\vskip 0.1in
Manuscript submitted on May 28, 2013, accepted on August 23, 2013
\begin{abstract}
A new criterion, based on noncontextuality, is derived to discriminate between separable and nonseparable states in classical wave optics where no discreteness is involved. An experiment is proposed to test the violation of noncontextuality by a nonseparable state. Such states have only recently begun to be explored. The significance of their nonseparability or entanglement as well as their similarities with and differences from entangled quantum states are discussed. 

\end{abstract}

\section{Introduction}
Entanglement has been held to be {\em the} trait of quantum mechanics that characterises its entire departure from classical physics \cite{schr}. Surprisingly, if non-factorizability or nonseparability of otherwise independent degrees of freedom of a system is taken as the main defining characteristic of `entanglement', then `entanglement' of the spatial and polarization degrees of freedom of classical light is an inevitable consequence of its Hilbert space structure, as was shown way back in 2001 \cite{ghose1,spreeuw}. Thus, classical optics and quantum mechanics share a common mathematical structure, which is after all not really surprising if classical optics is a limiting case of quantum optics. In fact, birefringent crystals have been known since the 17th century to produce states of the form $A u \otimes s + B v \otimes p$ where $u$ is the path with $s$ polarization and $v$ with $p$ polarization, $A$ and $B$ being amplitudes. Recently it has been shown that nonseparability of the spatial and polarization degrees of freedom of an inhomogeneously polarized light field (i.e. light fields whose polarization is not uniform over its spatial extension or support) is essential to provide the right physical basis to resolve the issue of the choice of the Mueller matrix \cite{simon}. Qian and Eberly have shown that classical light fields are intrinsically entangled in general, and that the degree of polarization of a field is the same as the degree of separation between the two spaces, complete separation occuring only for homogeneously polarized light beams \cite{eberly}. Cylindrically polarized laser beams which have been extensively studied and used in recent times \cite{oron, radial, holleczek,gabriel} are examples of this general result. Even Bell-like inequalities can be derived for classically entangled light beams, and their violation has been experimentally verified \cite{borges}. Interestingly, classical entanglement can be used to simulate many manipulations that are necessary for quantum information processing except, of course, those requiring nonlocality \cite{spreeuw}. 

Since `entanglement' in the sense of non-factorizability and nonseparability can and does occur in classical optics which is a paradigm of local theories, it is clear that entanglement {\em per se} and the Bell-CHSH inequalities that follow from it are neither specific to quantum mechanics nor do they imply nonlocality. Qian and Eberly have made the same point in a recent paper \cite{eberly2}. As mentioned towards the end of Ref. \cite{simon}, this aspect of entanglement is purely kinematic, arising, as it does, from the superposition principle of the tensor product of Hilbert spaces. Other aspects like nonlocality arise from the additional postulate of collapse or projective measurement and have no counterpart in classical optics. Furthermore, quantum mechanics operates on a Hilbert space of unit norm, which is not the case in classical optics where light beams can have arbitrary intensities. Consequently, not every quantum phenomena has a correspondence in classical optics.

In order to keep the distinction between the {\em implications} of entanglement in the classical and quantum cases in mind, various suggestions have been made about the correct nomenclature to be used in the classical case. Some have suggested the use of `nonseparability' \cite{borges}, some `structural inseparability' \cite{gabriel} and some `non-quantum entanglement' \cite{simon} in the case of classical optics. We would prefer to use `classical entanglement' to distinguish it from `quantum entanglement' but will also use nonseparability equivalently. The reason is to emphasize the fact that entanglement in the sense of nonseparability or non-factorizability is intrinsic to classical optics because of its Hilbert space structure which is not unique to quantum mechanics. 

A very important aspect of classical entanglement is its bearing on the concept of noncontextuality and realism. This has not been discussed at all so far. In this paper we wish to go into this question and show that the polarization and spatial modes of classically entangled light are {\em contextual} variables, which is a real surprise in classical physics.

In Section 2, a simple way to produce all four Bell-like states of polarization-path entangled classical light will be described. In Section 3 a new CHSH-Bell type of bound, implied by noncontextuality, will be derived for product states in classical wave optics without any discreteness assumption, and the entangled state $\vert \Phi^+)$ will be shown to violate this bound. The incompatibility of states like $\vert \Phi^-)$ with the axioms of the Kochen-Specker theorem will also be demonstrated. (We will use the notation $\vert X)$ and $(X\vert$ to distinguish classical states from quantum states $\vert X\rangle$ and $\langle X\vert$.) The important differences between classical and quantum entanglement will be summarised in Section 4. 

\section{Bell-like `States' in Classical Optics}

Classical electrodynamics is the paradigm of classical field theories in physics. That classical electrodynamics has a Hilbert space structure was first explicitly shown in 2001 \cite{ghose1}. Without getting involved in the details of that demonstration, let us simply note that two different Hilbert spaces are required for a complete description of an ordinary light beam in classical electrodynamics, namely a space $\hat{H}_{path}$ of square integrable functions that describe scalar optics, i.e. the spatial degrees of freedom of a classical light beam, and a two-dimensional space of polarization states $\hat{H}_{pol}$. These are disjoint Hilbert spaces, and hence a complete description of a classical light beam is given in terms of tensor products of `states' in these two Hilbert spaces (conventionally called modes): $\frac{1}{\sqrt{\vert A\vert^2}}\vert A)\otimes \vert \lambda)\in \hat{H}_{path}\otimes \hat{H}_{pol}$ where $A({\bf r},t) = ( {\bf r},t\vert A)$ are solutions of the scalar wave equation
\beq
\left[\nabla^2 - \frac{1}{c^2}\frac{\partial^2}{\partial t^2}\right]A({\bf r},t) = 0 
\eeq 
and $\vert \lambda)$ is the vector \[\vert \lambda) =\left(\begin{array}{c}
\lambda_1 \\\lambda_2
\end{array} \right) \] of the transverse polarizations $\lambda_1$ and $\lambda_2$. $A({\bf r},t)$ is a Laguerre-Gauss polynomial for laser beams. One can also write $\frac{1}{\sqrt{\vert A\vert^2}}\vert A)\otimes \vert \lambda)$ more conventionally as the Jones vector \[\vert J) = \frac{1}{\sqrt{( J\vert J)}}\left(\begin{array}{c}
 E_x\\ E_y
\end{array} \right) \] where $E_x = A_0 \hat{e}_x {\rm exp (i\phi_x)}$ and $E_y = A_0 \hat{e}_y {\rm exp (i\phi_y)}$ are the complex transverse electric fields, $\hat{e}_x$ and $\hat{e}_y$ are unit polarization vectors, and $\langle J\vert J\rangle = \vert E_x\vert^2 + \vert E_y\vert^2 =  A_0^2$ is the intensity $I_0$. Given this mathematical structure of a tensor product Hilbert space, polarization-path entanglement is inevitable because the tensor product space is also linear, allowing superposition of `states' in this space. Hence, it is possible to construct the complete set of Bell-like `states' in classical optics. 
\begin{figure}
{\includegraphics[scale=0.6]{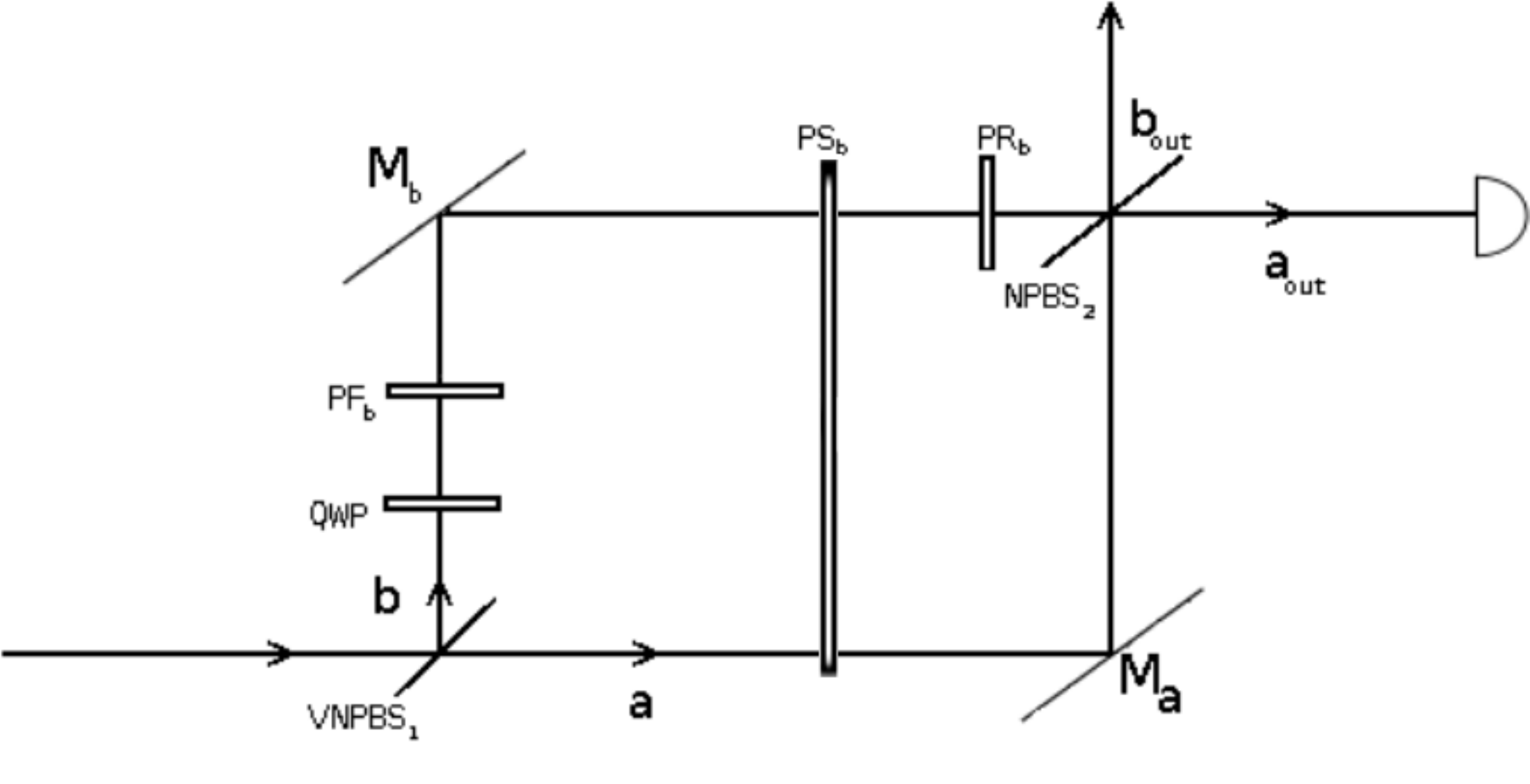}}
\caption{\label{Figure 1}{\footnotesize Schematic diagram of the experiment: A vertically polarized beam of classical light passes through a Mach-Zehnder interferometer. $PF_b$ in path $b$ is a polarization flipper that converts $V$ to $H$. $PS_b$ is a phase shifter and $PR_b$ is a polarization rotator in path $b$.}}
\end{figure}

Let us consider a classical light beam in a normalized mode $\vert \psi)_{in} = (A/\sqrt{I_0})\vert a_{in}) \otimes \vert V)$ of intensity $I_0 = \vert A \vert^2$ incident along the `path' $a_{in}$. When such a light beam is incident on a 50-50 lossless non-polarizing beam splitter $NPBS_1$ (Fig.1), the transmitted beam along the path $\vert a)$ and the reflected beam along the path $\vert b)$ span a two-dimensional sub-space $\hat{H}_{path}=\lbrace\vert a),\vert b)\rbrace$. Another two-dimensional Hilbert space $\hat{H}_{pol}=\lbrace\vert V),\vert H)\rbrace$ is associated with the polarization states of the two outgoing light beams. The action of $NPBS_1$ on the incident beam can be described by the unitary matrix 
\beq
U^{path}_{NPBS} = \frac{1}{\sqrt{2}}\left(\begin{array}{cc} 1 & \quad i \\ i & \,\,\,\,\,\,\,\,1\end{array}\right)_{path}\otimes \mathbb{I}^{pol}
\eeq
where $\mathbb{I}^{pol}$ is the unit operator on $\hat{H}_{pol}$, resulting in
\beq
U^{path}_{NPBS}\vert \psi)_{in} = \frac{A}{\sqrt{2I_0}}\left[\vert a)\otimes\vert V) + i \vert b)\otimes\vert V)\right],
\eeq which is a linear superposition of two coherent path modes of the same polarization and hence still a product `state'. When this is followed by a quarter-wave phase-shifter plate in path $b$ with its fast axis vertical, represented by the Jones matrix 
\beq
J_{QWP}= \left(\begin{array}{cc} 1 & \quad 0 \\ 0 & \,\,\,\,\,\,\,-i\end{array}\right)
\eeq
which acts on $\hat{H}_{pol}$, the unitary matrix $U$ that acts on the beam is
\beq
U = \vert a)( a\vert \otimes \mathbb{I}^{pol}+\vert b)( b\vert \otimes \hat{J}_{QWP},
\eeq
resulting in the output `state'
\ben
\vert\Psi_{out1}) &=&  \frac{A}{\sqrt{2I_0}}\left[\vert a)\otimes\vert V) + i \vert b)\otimes J_{QWP}\vert V)\right]\nonumber\\
&=& \frac{A}{\sqrt{2I_0}}\left[\vert a)\otimes\vert V) + \vert b)\otimes \vert V)\right].
\een
The action of the polarization flipper $PF_b$ on this `state' can be represented by the unitary matrix
\beq
U_{flip} = \vert a)( a\vert \otimes \mathbb{I}^{pol}+\vert b)( b\vert \otimes \hat{J}_{flip}
\eeq
where $\hat{J}_{flip}$ is the Jones matrix with $J_{xx} = J_{yy} = 0, J_{xy} = J_{yx} = 1$, so that $\hat{J}\vert V) = \vert H)$, and the final output `state' is
\beq
\vert\Phi^+) = U_{flip}\vert\Psi_{out1}) = \frac{A}{\sqrt{2I_0}}\left[\vert a)\otimes\vert V) + \vert b)\otimes\vert H)\right].\label{phi+}
\eeq
This is not a factorizable `state' but is polarization-path entangled, the initial single beam splitting into two separate path modes having different polarizations. One can introduce a $\pi$ phase shifter in the path $b$ to get the `state' $\vert \Phi^-)$. Similarly, one can get the other two Bell-like `states' $\vert \Psi^+)$ and $\vert \Psi^-)$ from a beam $\vert A) \otimes\vert H)$. The final entangled `state' after the second beam splitter $NPBS_2$ is
\beq
\vert\Phi^+)_f = \frac{A}{2\sqrt{I_0}}\left[\vert a)_{out}\otimes(i\vert V + \vert H)) + \vert b)_{out}\otimes(\vert V) + i \vert H)\right].
\eeq

The question that arises is: what kind of input classical light beam would be preferable to create a Bell-like `state' such as $\vert\Phi^+)$ or $\vert\Phi^-)$ on demand? Some discussion is necessary here. One difference between thermal and laser light is that the former is nondeterministic but the latter is not. Consequently, entanglement in thermal light is of a statistical character and uncontrollable. Ideal thermal light is a Bell-like state with zero polarization \cite{eberly2}. In the method proposed above, one needs to start with a product state of definite polarization. If thermal light is used, one has to insert a suitable polarization selector before the first beam splitter and destroy the statistical entanglement in the process. Once that is done, the nondeterministic character of the light plays no role in the rest of the required manipulations. Laser light, on the other hand, can be produced in a desired polarization mode. Furthermore, laser light is monochromatic and is preferable for interferometry over thermal light. Laser beams have already been used to create entanglement \cite{oron, radial, holleczek,gabriel}. Hence, laser light would be preferable. 

We emphasize that the word `state' in the classical context in which it is being used in this paper does not imply any quantization and Fock states in quantum field theory. With this understanding, we will henceforth drop the quotes from the word `state'.

\section{Noncontextuality: A New Bound}
Our next aim is to test whether these classically entangled states are consistent with noncontextuality, namely the conventional {\em classical} notion that a physical property must be independent of the context in which it is measured. It has always been a tenet of classical physical science that whatever exists in the physical world is independent of our observations which only serve to reveal them to us.  Put more technically, what this means is that the result of a measurement is predetermined and is not affected by how the value is measured, i.e. not affected by previous or simultaneous measurement of any other compatible or co-measureable observable. Local theories are particular examples of noncontextuality, because the result of a measurement in such theories does not depend on measurements made simultaneously on spatially separated (mutually non-interacting) systems. Noncontextuality of hidden variables introduced in interpretations of quantum mechanics is a straightforward generalization of this notion \cite{ks,mermin,noncontext}. To test this classical concept, joint measurements of compatible observables that are not necessarily spatially separated are required. This can be done for classical light by making a joint measurement of its two otherwise independent degrees of freedom, namely path and polarization, in one path. It will be shown that the requirement of noncontextuality in such a case results in a CHSH-like bound on correlations formally similar to the one derivable in the quantum mechanics of a single particle whose path and spin are entangled \cite{home}, but with the difference that no discreteness assumption is made. If the light is prepared in a non-factorizable state like $\vert\Phi^+)$, it is possible to check if the predicted correlations satisfy this CHSH-like bound. It must be remembered that {\em it is not necessary to invoke any hidden variables in this classical case} simply because the theory itself has always stood the tests of locality and determinism, and until now its noncontextuality has never been challenged. 

All derivations of Bell-CHSH inequalities to date have been based on the assumption of local realism and the discreteness of outcomes of measurement. Correlated pairs of {\em particles} are taken to move such that one enters one apparatus and the other a distant apparatus. In each apparatus the particle selects one of two channels labeled $+1$ and $-1$. Note that this discreteness is not purely quantum in nature but occurs in all particle theories. In fact, Bell's theorem is not derived from quantum mechanics at all--it is a bound that {\em all} theories satisfying local-realism must satisfy. Quantum mechanics, in fact, is incompatible with this bound. To test whether or not Bell's theorem is violated by quantum systems, it is therefore necessary to invoke strong projective measurements in quantum theory to ensure discreteness in the outcomes of measurement.

This discreteness is absent in classical optics where intensities of light are measured, and these can vary continuously. Nevertheless, since path-polarization entanglement occurs in classical optics, it is important to check whether noncontextuality holds in such cases. One must therefore first derive a bound similar to the CHSH-Bell bound that noncontextuality implies for product states in classical optics. This is what we proceed to do now.

Let us define a correlation
\begin{eqnarray}
E(\theta,\phi)=(\Psi\vert \sigma_{\theta}.\sigma_{\phi}\vert\Psi)
\end{eqnarray}
where $\vert \Psi)$ is an arbitrary normalized optical state and
\begin{eqnarray}
\sigma_{\theta}=\sigma_{\theta,0}-\sigma_{\theta,\pi},\\
\sigma_{\phi}=\sigma_{\phi,0}-\sigma_{\phi,\pi},
\end{eqnarray}
with
\begin{eqnarray}
\sigma_{\theta,0}=\frac{1}{2}(|V)+e^{i\theta}|H))((V|+e^{-i\theta}( H|)\otimes I_{path},\nonumber \\
\sigma_{\theta,\pi}=\frac{1}{2}(|V)-e^{i\theta}|H))(( V|-e^{-i\theta}( H|)\otimes I_{path},\nonumber \\
\sigma_{\phi,0}=I_{pol}\otimes\frac{1}{2}(|a)+e^{i\phi}|b))(( a|+e^{-i\phi}( b|),\nonumber \\
\sigma_{\phi,\pi}=I_{pol}\otimes\frac{1}{2}(|a)-e^{i\phi}|b))(( a|-e^{-i\phi}( b|).\label{sigma}
\end{eqnarray}
Hence,
\begin{eqnarray}
\sigma_{\theta}=(e^{-i\theta}|V)( H|+e^{i\theta}|H)( V|)\otimes I_{path},\\
\sigma_{\phi}=I_{pol}\otimes (e^{-i\phi}|a)( b|+e^{i\phi}|b)( a|).
\end{eqnarray}
These projection operators represent polarization and path measurements in classical optics. It should be noted that $\sigma_{\theta}$ and $\sigma_{\phi}$ act upon different Hilbert spaces altogether, one belonging to path and the other belonging to polarization. Hence they commute with each other.

Now, a general normalized product state can be written as 
\begin{eqnarray}
|\Psi)=|\psi_{pol})|\psi_{path})=(\cos\alpha|V)+e^{i\beta}\sin\alpha|H))(\cos\gamma|a)+e^{i\delta}\sin\gamma|b))
\end{eqnarray}
where $\alpha, \beta, \gamma, \delta$ are arbitrary parameters.
For such a state
\begin{eqnarray}
E(\theta,\phi)&=&(\psi_{pol}|\sigma_{\theta}|\psi_{pol})(\psi_{path}|\sigma_{\phi}|\psi_{path})\nonumber \\
&=& E_{pol}(\theta)E_{path}(\phi),\label{exp}
\end{eqnarray}
with
\begin{eqnarray}
E_{pol}(\theta)=\sin\alpha\cos(\beta-\theta),\label{epol}\\
E_{path}(\phi)=\sin\gamma\cos(\delta-\phi).\label{epath}
\end{eqnarray}
Thus, the expectation value $E(\theta,\phi)$ is the product of the expectation values of the polarization and path. Hence, the path and polarization measurements for product states in classical optics are independent of one another in all contexts. This is the content of noncontextuality. This may, at first sight, look obvious and trivial, but on closer inspection, one finds that it implies the inequality
\beq
-1\leqslant E(\theta,\phi)\leqslant 1
\eeq
for the correlation.

Now, define a quantity $S$ as
\ben
S(\theta_{1},\phi_{1};\theta_{2},\phi_{2}) &=& E(\theta_{1},\phi_{1})+E(\theta_{1},\phi_{2})-E(\theta_{2},\phi_{1})+E(\theta_{2},\phi_{2})\nonumber\\
&\equiv& m(b+a)+n(b-a)
\een
with
\begin{eqnarray}
m=E_{pol}(\theta_{1}),\,\,  n=E_{pol}(\theta_{2}),\nonumber \\
a=E_{path}(\phi_{1}),\,\,   b=E_{path}(\phi_{2}).
\end{eqnarray}
Using the triangle inequality 
\begin{eqnarray}
|S|\leq |m||b+a|+|n||b-a|
\end{eqnarray}
and remembering that $|m|,|n|\leq 1$, it follows that
\begin{eqnarray}
|S|\leq |m||b+a|+|n||b-a|\leq |b+a|+|b-a|.
\end{eqnarray}
Since
\begin{eqnarray}
{\rm max}(A,B)=\frac{A+B}{2}+\frac{|A-B|}{2},
\end{eqnarray}
it follows that
\beq
|S|\leq 2\label{bound}
\eeq
for all possible combinations of values of $a,\,b \leq \pm 1$.
All that is required to derive this bound for product states is that {\em the correlations lie between $-1$ and $+1$}, which is guaranteed by the results (\ref{epol}) and (\ref{epath}). Unlike in the usual Bell case for particle mechanics, no discreteness assumption is necessary. This is therefore a new and non-trivial result. It shows that  the form of CHSH-Bell inequalities do not necessarily imply any discreteness.

Now consider the correlation calculated for the normalized state (\ref{phi+}) given by
\ben
E(\theta,\phi)&=&(\Phi^+|\sigma_{\theta}\cdot\sigma_{\phi}|\Phi^+)\nonumber\\
&&\:=( \Phi^+\vert \,[(+)\sigma_{\theta,0} + (-)\sigma_{\theta,\pi} ]. [(+)\sigma_{\phi,0} + (-)\sigma_{\phi,\pi} ]\vert\Phi^+).
\een
This correlation can be measured by measuring the intensities of light at the final detector in path $a_{out}$ (Fig. 1) corresponding to four possible combinations of orientation of the phase-shifter $PS_b$ and the polarization rotator $PR_b$ in path b as follows:
\ben
E(\theta, \phi) &=& (\Phi^+|[\sigma_{\theta,0}\cdot\sigma_{\phi,0}+\sigma_{\theta,\pi}\cdot\sigma_{\phi,\pi}\nonumber\\ &&\: -\sigma_{\theta,0}\cdot\sigma_{\phi,\pi}-\sigma_{\theta,\pi}\cdot\sigma_{\phi,0}]|\Phi^+).\label{E} 
\een
The intensities corresponding to the four possible orientations are given by
\ben
I(\theta,\phi)&=&( \Phi^+|\sigma_{\theta,0}\cdot\sigma_{\phi,0}|\Phi^+),\nonumber\\
I(\theta+\pi,\phi+\pi)&=&( \Phi^+|\sigma_{\theta,\pi}\cdot\sigma_{\phi,\pi}|\Phi^+),\nonumber\\
I(\theta+\pi,\phi)&=&( \Phi^+|\sigma_{\theta,\pi}\cdot\sigma_{\phi,0}|\Phi^+),\nonumber\\
I(\theta,\phi+\pi)&=&( \Phi^+|\sigma_{\theta,0}\cdot\sigma_{\phi,\pi}|\Phi^+),\nonumber\\
\een
where clearly $I(\theta,\phi)=\frac{1}{2}[1+\cos(\theta+\phi)]$ from (\ref{phi+}) and the definitions (\ref{sigma}). This sinusodial behavior of the intensity is consistent with classical wave interference.
We can write $E(\theta, \phi)$ in terms of the normalized intensities as
\ben
E(\theta,\phi)&=&\dfrac{I(\theta,\phi)+I(\theta+\pi,\phi+\pi)-I(\theta+\pi,\phi)-I(\theta,\phi+\pi)}{I(\theta,\phi)+I(\theta+\pi,\phi+\pi)+I(\theta+\pi,\phi)+I(\theta,\phi+\pi)}\nonumber\\
 &=& \cos(\theta+\phi).
\een
This is not in a product form like (\ref{exp}). It is also clear from this that the noncontextuality bound (\ref{bound}) is violated by the state $\vert \Phi^+)$ for the set $\theta_1 = 0, \theta_2 = \pi/2, \phi_1 = \pi/4, \phi_2 = - \pi/4$ for which $\vert S\vert = 2\sqrt{2}$. This violation shows that the path and polarization of even classical light in entangled states like $\vert \Phi^+)$ are contextual. {\em Since the path and polarization changes are made on the same state in path $b$, there is no violation of locality in this result}. 

One can further show that classical entangled states can violate the axioms of the Kochen-Specker theorem. Following Mermin and Peres \cite{mermin2}, consider the state
\beq
\vert\Phi^-) = \frac{A}{\sqrt{2I_0}}\left[\vert a)\otimes\vert V) - \vert b)\otimes\vert H)\right]
\eeq
and six hermitian operators $J_x^{pol}, J_x^p, J_y^{pol}, J_y^p, J_x^{pol} J_y^p, J_y^{pol} J_x^p$  acting on it, where $J^{pol}$s are Jones matrices acting on $\hat{H}_{pol}$ and $J^p$s are analogous matrices acting on $\hat{H}_{path}$. Hence, $J^{pol}$s and $J^p$s commute. All the operators mutually commute except the last two. As is well known, these matrices may be taken to have the same form as the Pauli matrices $\sigma_i$ with the property $\sigma_x\sigma_y = i\sigma_z$. They have eigenvalues $\pm 1$.
Since 
\ben
J_x^p\vert a) &=& \vert b),\,\,\,\,\,\, J_x^p\vert b) = \vert a),\\
J_y^p \vert a) &=& i\vert b),\,\,\,\,\,\, J_y^p \vert b) = -i\vert a),\\
J_x^{pol}\vert V) &=& \vert H),\,\,\,\,\,\, J_x^{pol}\vert H) = \vert V),\\
J_y^{pol}\vert V) &=& -i\vert H),\,\,\,\,\,\, J_y^{pol}\vert H) = i\vert V),
\een
it follows that
\ben
J_x^{pol}\,.\, J_x^p \vert\Phi^-) &=& -\vert\Phi^-),\\
J_y^{pol}\,.\, J_y^p \vert\Phi^-) &=& -\vert\Phi^-),\\
J_x^{pol} J_y^p\, .\, J_x^{pol}\,.\,J_y^p \vert\Phi^-) &=& +\vert\Phi^-),\\
J_y^{pol} J_x^p\, .\, J_y^{pol}\,.\,J_x^p \vert\Phi^-) &=& +\vert\Phi^-),\\
J_x^{pol} J_y^p\, .\, J_y^{pol}J_x^p \vert\Phi^-) &=& -\vert\Phi^-).
\een
Hence, the state $\vert\Phi^-)$ is an eigenstate of the operators on the left-hand sides with eigenvalues $\pm 1$.
Since each of the six operators $J_x^{pol},J_x^p,J_y^{pol},J_y^p,\\J_x^{pol} J_y^p,J_y^{pol} J_x^p$ occurs exactly twice on the left-hand sides, the product of the left-hand sides is $+1$ provided each property of the state has a predetermined and context independent value $\pm 1$. However, the product of the right-hand sides is $-1$. This is a clear logical contradiction. The Kochen-Specker theorem is based on two assumptions, namely (i) value definiteness, i.e. all physical properties or observables of a system have predetermined values, and (ii) noncontextuality, which requires these values to be independent of the way in which they are measured \cite{held}. Hence, these assumptions cannot hold for entangled classical states like $\vert\Phi^-)$.

\section{Implications and Significance}

Although there is no doubt that quantum entanglement has more implications than those of classical entanglement because of {\em additional assumptions} like the use of projective spaces and projective measurement, it is clear from the foregoing discussions that there is more to classical electrodynamics than meets the eye. Classical electrodynamics and quantum mechanics share a Hilbert space structure which gives rise to many common features like entanglement and the violation of Bell-CHSH-like inequalities which are novel features of classical optics that have only recently begun to be explored. We have derived an important criterion or bound in classical wave optics which shows that although the path and polarization of separable states are noncontextual variables, those of nonseparable states like $\vert\Phi^+)$ are not. 

The superposition principle resulting in interference phenomena in classical optics is a straightforward consequence of its Hilbert space structure. There is a mathematical theorem which states that {\em every pair of vector spaces has a tensor product} \cite{stern}. The tensor product space is also a linear vector space. States of classical light are tensor products of two linear vector spaces, the Hilbert space of space-time functions (scalar optics) and the Hilbert space of polarization. Hence, {\em the existence of tensor product spaces resulting in polarization-path entanglement is just as inevitable in classical optics as in quantum mechanics}. It should be clear from this that classical optics (electrodynamics) is a lot more like quantum mechanics than is classical mechanics in which states are points in phase space. 

What are the additional assumptions in quantum mechanics that differentiate it from classical optics? One such difference is that physical states in quantum mechanics are of unit norm and therefore lie on a unit sphere in Hilbert space. This is necessary for the probabilistic interpretation. Also, all states that differ only by an overall phase factor are identified, and hence quantum nechanics actually operates on coset spaces. This is not the case with classical optics which allows beams of light of arbitrary intensities, the fluctuations being of purely classically statistical in character. The second crucial difference is the postulate of strong projective measurement. All this, for example, leads to a significant difference in the interpretation of superposition of states in quantum mechanics and classical optics. In quantum mechanics a physical state in general does not {\em possess} physical properties {\em before} measurement. For example, consider a single-photon state $\vert X\rangle = c_1\vert V\rangle + c_2 \vert H\rangle$ with $\vert c_1\vert^2 + \vert c_2\vert^2 = 1$. On measurement in the basis $(V, H)$, the state is projected to either $\vert V\rangle$ with probability $\vert c_1\vert^2$ or $\vert H\rangle$ with probability $\vert c_2\vert^2$. This is not the case in classical optics where an analogous measurement will always give the two classical amplitudes $c_1$ and $c_2$ simultaneously. Hence, it can be said to possesses a definite polarization. This {\em innocuous scientific realism} of the polarization states in classical optics, however, holds only for product states like $(c_1 \vert x\rangle + c_2 \vert y\rangle)\otimes \vert w\rangle$ but not for a superposition of product states like $c_1 \vert H\rangle \otimes \vert x\rangle + c_2 \vert V\rangle \otimes \vert y\rangle$ in which the polarization and path degrees of freedom are entangled. Such states cannot be said to possess either a definite path or a definite polarization, before or after a measurement. Hence, these states are of fundamental significance in classical optics, though they have not been considered possible until recently. They encode correlations between polarization states and path states similar to path-spin entangled states in quantum mechanics. They therefore violate a CHSH-Bell-like inequality derived from the requirement of noncontextuality, as shown in Section 2 for $\vert \Phi^+)$. States like $\vert \Phi^-)$ also violate the axioms of the Kochen-Specker theorem. Hence, polarization and path are {\em contextual} degrees of freedom for such classical light. They do not, however, imply any nonlocality. Hence, noncontextuality can be violated without violating locality.

Since physicists became familiar with the term noncontextuality from the classic work of Kochen and Specker, it is invariably associated in their minds with hidden variable theories, and hence its use in classical optics where there are no hidden varibles appears paradoxical. As is clear from the classic paper, the concept of noncontextuality is entirely classical and does not hold in general in quantum mechanics. That it does not hold generally in classical optics as well is the very surprising and significant result.

In spite of the differences, the similarity between quantum optics and classical uptics is striking. {\em Innocuous classical realism, i.e. separability and noncontextuality, are as impossible to reconcile with quantum mechanics as with classical optics}. To hold on to realism in some form in classical physics as a whole, one has therefore to search for a deeper and more subtle meaning of reality than is captured by separability and noncontextuality.

\section{Acknowledgement}
We thank C. S. Unnikrishnan, Dipankar Home, Arun Pati, Ujjwal Sen and Tony Sudbery for helpful comments on the preliminary drafts of the paper. The views expressed in this paper are, however, entirely ours. PG thanks the National Academy of Sciences, India for the award of a Senior Scientist Platinum Jubilee Fellowship which allowed this work to be undertaken.

\end{document}